\documentclass[
aps,
pra,
onecolumn,
twocolumn,
showpacs,
superscriptaddress,
floatfix
]{revtex4-1}
\usepackage[dvipdfmx]{graphicx}
\usepackage{verbatim}
\usepackage{epsf}
\usepackage{natbib}
\usepackage{dcolumn}
\usepackage{bm}
\usepackage{color} 
\usepackage{ulem}
\usepackage{amsmath}
\usepackage{amssymb}
\usepackage{txfonts}
\usepackage{adjustbox}
\usepackage{tikz}
\usetikzlibrary{matrix,arrows,decorations.pathmorphing}
\usepackage[
colorlinks=true,
citecolor=blue,
urlcolor=blue,
setpagesize=false]{hyperref}
\newcommand{\dd}   {\mathrm{d}}
\newcommand{\e}    {\mathrm{e}}
\newcommand{\up}   {\uparrow}
\newcommand{\dn}   {\downarrow}

\newcommand{\s}    {\sigma}

\hypersetup{
  colorlinks=true,
  linkcolor=[rgb]{0.60,0.00,0.0},
  citecolor=[rgb]{0.0,0.0,0.6},
  urlcolor=[rgb]{0.0,0.0,0.6},
  setpagesize=false
}
\begin{document}

\title{
  Benchmark study of an auxiliary-field quantum Monte Carlo technique
  for the Hubbard model with shifted-discrete Hubbard-Stratonovich transformations
}

\author{Kazuhiro~Seki}
\affiliation{Scuola Internazionale Superiore di Studi Avanzati, via Bonomea, 265-34136 Trieste, Italy}
\affiliation{Computational Materials Science Research Team, RIKEN Center for Computational Science (R-CCS), Kobe, Hyogo 650-0047,  Japan}
\affiliation{Computational Condensed Matter Physics Laboratory, RIKEN, Wako, Saitama 351-0198, Japan}

\author{Sandro~Sorella}
\affiliation{Scuola Internazionale Superiore di Studi Avanzati, via Bonomea, 265-34136 Trieste, Italy}
\affiliation{Computational Materials Science Research Team, RIKEN Center for Computational Science (R-CCS), Kobe, Hyogo 650-0047,  Japan}
\affiliation{Democritos Simulation Center CNR--IOM Istituto Officina dei Materiali, Via Bonomea 265, 34136 Trieste, Italy}            

\begin{abstract} 
  Within  the ground-state auxiliary-field quantum Monte Carlo technique,  
  we introduce discrete Hubbard-Stratonovich transformations (HSTs)
   that are suitable also  for spatially inhomogeneous trial functions. 
  The discrete auxiliary fields introduced here are 
  coupled to local spin or charge operators fluctuating around 
  their Hartree-Fock values. 
  The formalism can be considered as
  a generalization of the discrete HSTs by Hirsch~ 
  \href{https://link.aps.org/doi/10.1103/PhysRevB.28.4059}
       {[J. E. Hirsch, Phys. Rev. B {\bf 28}, 4059 (1983)]}
  or a compactification of the shifted-contour auxiliary-field Monte Carlo formalism by Rom {\it et al.}~
  \href{http://www.sciencedirect.com/science/article/pii/S0009261497003709}
       {[N. Rom {\it et al.}, Chem. Phys. Lett. {\bf 270}, 382 (1997)]}.   
  An improvement of the acceptance ratio is found for a real auxiliary field, while
  an improvement of the average sign is found for a pure-imaginary auxiliary field. 
  Efficiencies of the different HSTs are tested in the single-band Hubbard model 
  at and away from half filling by studying the staggered magnetization
  and energy expectation values, respectively.   
\end{abstract}

\date{\today}

\maketitle
\section{Introduction}

The numerical solution of the Hubbard model 
with strong correlations is one of the most challenging issues in 
the theory of strongly correlated electron systems~\cite{LeBlanc2015,Zheng2017}. 
Attempts to determine  the ground state are often
based on iterative techniques based 
on a repeated application of a short imaginary time propagator, or by using 
the simple power method and  
more advanced 
Krylov-subspace techniques, as, for instance the Lanczos algorithm, 
where a Hamiltonian operator is repeatedly applied 
to a properly chosen trial state. 
In both cases the ground-state component of the trial state is filtered out after several iterations.

Among these projection techniques, the  
auxiliary-field quantum Monte Carlo (AFQMC)~\cite{Sugiyama1986,Sorella1989,Imada1989,Becca_Sorella_book}
is one of the most powerful schemes, as it allows us to study,
for example, the ground-state properties of the Hubbard model 
with several thousands electrons and lattice sites,  
when the negative-sign problem is absent~\cite{Sorella2012,Otsuka2016,Otsuka2018}. 
In the ground-state AFQMC, even if the Hamiltonian is the same, 
there exists some  arbitrariness in choosing  
the  trial wave function and the type of auxiliary fields (e.g., real, complex, continuous, or discrete). 
Experience has shown that an appropriate choice of these ingredients may significantly improve the efficiency of the 
Monte Carlo simulations~\cite{Shi2013}.

It has been demonstrated~\cite{Qin2016,Zheng2017} 
that a Slater-determinant obtained from 
an unrestricted Hartree-Fock (UHF) approximation~\cite{Xu2011} 
provides  a good trial wave function 
for the doped Hubbard model in the constrained-path AFQMC~\cite{Zhang1997}. 
Recently, 
for a particular parameter set at doping $\delta=1/8$ 
and electron-electron repulsion $U/t=8$, 
the ground state of the Hubbard model on the square lattice 
has been predicted to exhibit a vertical stripe order~\cite{Zheng2017},
where the stripe states with periods $\lambda=5,6,7$ and $8$ 
in units of the lattice constant
are nearly degenerate,
while a spatially homogeneous $d$-wave superconducting state 
should  have, according to their study,  a higher energy. 
Recent variational Monte Carlo (VMC) calculations~\cite{Zhao2017,Ido2018,Darmawan2018} have also 
shown that various vertical-stripe orders  
with different periods appear 
depending on the doping and the hopping parameter. 
In most of the calculations in Ref.~\cite{Zheng2017},
the symmetry of finite-size clusters 
is broken due to the use of UHF trial wave functions or
by applying pinning magnetic fields, 
and the results are extrapolated to the thermodynamic limit. 
The success of utilizing symmetry-broken wave functions is rather surprising, 
because symmetry breakings do not occur in 
the exact ground state of finite-size systems. 
The similar issue is known as the symmetry dilemma 
in first-principles calculations for molecules~\cite{Perdew1995,Carrascal2015}. 
Recently, it has been shown that the 
quality of the trial wave function can be improved 
by restoring the symmetries that are once broken by UHF 
or mean-field treatments~\cite{Tahara2008,Rodriguez2012,Shi2014}.  
However, in the present work, we do not enter into
the issue on symmetry breakings of 
trial wave functions, and rather focus on
the arbitrariness of the auxiliary field 
to improve the efficiency of AFQMC simulations with such 
symmetry-broken trial wave functions.

The way of transforming a quartic interaction term into a quadratic one via 
the Hubbard-Stratonovich transformation (HST)~\cite{Hubbard1959} 
is not unique and affects the efficiency of simulations~\cite{Hirsch1983,Motome1997,Held1998,Sakai2004,Han2004,Broecker2016}.
Recently the popularity of this technique is substantially increased, because it has been realized that, 
with continuous auxiliary fields, one can 
treat interaction terms beyond the on-site Hubbard interaction, up to the complete treatment of the  
long-range Coulomb interaction~\cite{Buividovich2012,Ulybyshev2013,Hohenadler2014,Tang2015,Tang2018}, 
 or of the long-range electron-phonon interaction~\cite{Batrouni2019}, and 
even both of them on the same footing~\cite{Karakuzu2018}, 
without being vexed by the sign problem in
a certain parameter region on bipartitle lattices. 
Interestingly, such a parameter region coincides with the one where
rigorous statements on the ground state of an extended Hubbard-Holstein
model are available~\cite{Lieb1989,Miyao2018}. 
It is also noteworthy that, even when the sign problem cannot
be eliminated completely, continuous auxiliary fields
with a proper shift~\cite{Rom1997,Rom1998,Baer1998} 
can improve the efficiency of simulations
compared to the one without the shift. 
A similar idea has been employed also in the AFQMC~\cite{Zhang2003,Motta2014}  
within the constrained-path approximation~\cite{Zhang1997}. 

In this paper, we introduce shifted-discrete HSTs,  
where auxiliary fields are coupled to the fluctuation of local spin or charge. 
The method is applied for AFQMC simulations of the Hubbard model on the square lattice. 
It is shown that the shifted-discrete HSTs can improve the efficiency of
the AFQMC simulations. 
Moreover we present results on the magnetic order parameter as a function of $U/t$ with high statistical 
accuracy, that represents important benchmark, useful also for comparison with experiments.

The rest of the paper is organized as follows. 
In Sec.~\ref{sec:2}, the Hubbard model is defined and the AFQMC method is described.
In Sec.~\ref{sec:3}, the shifted-discrete HSTs are introduced. 
In Sec.~\ref{sec:4}, numerical results of the AFQMC simulations 
for the Hubbard model are presented. 
Section~\ref{sec:5} is devoted to conclusions and discussions.

\section{Model and method}
\label{sec:2}
We consider the Hubbard model whose Hamiltonian is defined by 
$\hat{H} = \hat{K}  + \hat{V}$, 
where 
\begin{eqnarray}
  \label{Ham0}
  \hat{K} &=& -t\sum_{\langle ij \rangle, \s} 
  \left( \hat{c}_{i\s}^\dag \hat{c}_{j\s} + {\rm H.c.} \right), \\
  \hat{V} &=& U\sum_{i} \hat{n}_{i \up} \hat{n}_{i \dn},  \label{HamU}
\end{eqnarray}
$\hat{c}_{i \s}^\dag$ ($\hat{c}_{i \s}$) creates (annihilates) a 
fermion with site index $i$ and spin index $\s(=\up,\dn)$, 
$\hat{n}_{i\s} = \hat{c}_{i\s}^\dag \hat{c}_{i\s}$, 
$t$ is the hopping parameter between the nearest-neighbor sites on the square lattice 
and $U > 0$ is the on-site electron-electron repulsion. 
We consider the Hubbard model on 
$N=L \times L$-site clusters.
Boundary conditions will be specified for each calculation in Sec.~\ref{sec:4}. 
The lattice constant is set to be unity.

In the AFQMC, the expectation value of 
an operator $\hat{O}$ is calculated as 
\begin{equation}
  \langle \hat{O} \rangle_{\beta} = 
  \frac{
    \langle \Psi_{\rm T} |
    \e^{-\frac{\beta}{2} \hat{H}} \hat{O} 
    \e^{-\frac{\beta}{2} \hat{H}}| \Psi_{\rm T} \rangle }
       {\langle \Psi_{\rm T} | \e^{-\beta \hat{H}}  |\Psi_{\rm T} \rangle }, 
       \label{afqmc}
\end{equation}
where $\beta$ is the projection time 
and $ |\Psi_{\rm T} \rangle $ is a 
trial wave function.
If $\beta$ is infinitely large, one can obtain the ground-state
expectation value as long as $|\Psi_{\rm T} \rangle$ has a finite
overlap with the ground state~\cite{Horn1984}. 
If $\beta$ is finite, the results depend on the trial wave function
(see for example Ref.~\cite{Weinberg2017}).
If $\beta=0$, Eq.~(\ref{afqmc}) reduces to the expectation
value of $\hat{O}$ with respect to the trial wave function.

At finite dopings, 
$|\Psi_{\rm T} \rangle$ is obtained by solving the eigenvalue problem of the 
following UHF Hamiltonian self consistently: 
\begin{equation}
\hat{H}_{\rm UHF} = \hat{K}
+ U_{\rm eff} \sum_{i}\left(
\langle \hat{n}_{i\up}  \rangle_0 \hat{n}_{i\dn}
+ \hat{n}_{i\up}  \langle \hat{n}_{i\dn} \rangle_0
-
\langle \hat{n}_{i\up} \rangle_0
\langle \hat{n}_{i\dn} \rangle_0
\right),
\label{eq:uhf}
\end{equation}
where $U_{\rm eff}$ is an arbitrary parameter and 
the expectation value $\langle \cdots \rangle_0$ in Eq.~(\ref{eq:uhf})
is defined in Eq.~(\ref{afqmc}) with $\beta=0$. 
A fine tuning of $U_{\rm eff}$ 
can improve the quality of the trial wave function~\cite{Qin2016}.  
We set $U_{\rm eff}/t=2.5$ 
which has turned out to provide a good trial wave function 
for the doped cases studied here, 
in the sense that the energy expectation value 
decreases quickly with increasing $\beta$.  
By adding a small bias in the initial condition 
for the self consistent UHF loop to pin the direction of the stripe, 
$|\Psi_{\rm T} \rangle$ shows a vertical stripe order 
with period $\lambda=8$ around $\delta=1/8$ doping on the $16 \times 16$ cluster.

At half filling, $|\Psi_{\rm T} \rangle$ is obtained as a ground state 
of non-interacting electrons on the square lattice under
a staggered magnetic field along the spin-quantized axis ($z$ direction): 
\begin{equation} 
  \hat{H}_{\rm MF} = \hat{K} - \Delta_{\rm AF} \sum_{i} (-1)^{i} (\hat{n}_{i\up} - \hat{n}_{i\dn}),
  \label{eq:Hmf}
\end{equation}
where $(-1)^i = 1 (-1)$ if the site $i$ belongs to the $A$ ($B$) sublattice
and $\Delta_{\rm AF}$ can be chosen arbitrarily. 
The value of $\Delta_{\rm AF}$ will be specified 
with the numerical results in Sec.~\ref{sec:4}.

By using the second-order Suzuki-Trotter decomposition~\cite{Trotter1959,Suzuki1976},
the imaginary-time propagator can be expressed as 
\begin{equation}
  \e^{-\beta \hat{H}} = \prod_{n=1}^{N_\tau} 
  \left(
  \e^{-\frac{\Delta_\tau}{2} \hat{K}} 
  \e^{-\Delta_\tau \hat{V}} 
  \e^{-\frac{\Delta_\tau}{2} \hat{K}} 
  \right) + O\left(\Delta_\tau^2\right), 
  \label{Trotter}
\end{equation}
where the projection time $\beta$ is discretized into $N_\tau$ time slices 
and $\Delta_\tau=\beta/N_\tau$.
For the doped cases, we set $\Delta_\tau t=0.05$ 
so that the discretization error is within statistical errors. 
For the half filled case,
we perform extrapolations of $\Delta_\tau \to 0$ to eliminate the discretization error,  
which becomes non negligible for large $U/t$ 
as compared to statistical or extrapolation
errors for the results shown in Sec.~\ref{sec:result_half}. 
An HST is applied to $\e^{-\Delta_\tau \hat{V}}$ and 
the summation over the auxiliary fields is performed 
by the Monte Carlo method with the importance sampling,
where a proposed auxiliary-field configuration is accepted or rejected
according to the Metropolis algorithm. 
In the next Section, we introduce  
shifted-discrete HSTs for $\e^{-\Delta_\tau \hat{V}}$.

\section{Shifted-discrete Hubbard-Stratonovich transformations}\label{sec:3}
In this Section we derive shifted-discrete HSTs
which couple the auxiliary field
to the local spin fluctuation in Sec.~\ref{hst:spin} and
to the local charge fluctuation in Sec.~\ref{hst:charge}. 
Although the two HSTs can be formulated almost in parallel, 
we provide both of them separately for completeness.

\subsection{Auxiliary field coupled to spin fluctuation}\label{hst:spin}
The Hubbard interaction in Eq.~(\ref{HamU}) can be written as 
\begin{eqnarray}
  \hat{V} = 
  &-&\frac{U}{2} \sum_{i} 
  \left[\left(\hat{n}_{i\up} -\hat{n}_{i\dn} - \tilde{m}_{i} \right)^2 
    -\tilde{m}_{i}^2\right]\notag \\
  &+&\frac{U}{2} \sum_{i} \left[ 
    (1-2 \tilde{m}_{i}) \hat{n}_{i\up} +
    (1+2 \tilde{m}_{i}) \hat{n}_{i\dn} \right]
  \label{eq:HamUspin}
\end{eqnarray}
where $\tilde{m}_i$ is an arbitrary number.  
Then $\e^{-\Delta_\tau V}$ can be written as 
\begin{eqnarray}
  \e^{-\Delta_\tau \hat{V}} &=&
  \e^{ \frac{\Delta_\tau U}{2}\sum_{i} \left[(\hat{n}_{i\up} - \hat{n}_{i\dn} - \tilde{m}_{i})^2 - \tilde{m}_{i}^2\right]} \notag \\
  &\times&
  \e^{-\frac{\Delta_\tau U}{2}\sum_{i} (1-2\tilde{m}_i)\hat{n}_{i\up}}
  \e^{-\frac{\Delta_\tau U}{2}\sum_{i} (1+2\tilde{m}_i)\hat{n}_{i\dn}}.  
  \label{propUspin}
\end{eqnarray}
Let us consider the first exponential factor in 
the right-hand side of Eq.~(\ref{propUspin}). 
For each site $i$, we consider the following HST:  
\begin{equation}
  C_i \e^{-\frac{\Delta_\tau U}{2}\tilde{m}_i^2} \e^{\frac{\Delta_\tau U}{2}(\hat{n}_{i\up} - \hat{n}_{i\dn} - \tilde{m}_i)^2} 
  = \frac{1}{2} \sum_{s_i=\pm1} \e^{ \alpha_i s_i (\hat{n}_{i\up} - \hat{n}_{i\dn} - m_{i})}, 
  \label{assumption}
\end{equation}
where $s_i=\pm 1$ is the discrete auxiliary field,
and the undetermined four parameters $\alpha_i$, $m_i$, $\tilde{m}_i$ and $C_i$ 
are related through the following three equations 
(see Appendix~\ref{appA} for derivation): 
\begin{eqnarray} 
  &&\frac{\cosh{\alpha_i (1-m_i)} \cosh{\alpha_i(1+m_i)}}{\cosh^2{\alpha_i m_i}}= \e^{\Delta_\tau U}, \label{alpha_spin}\\
  &&\tilde{m_i} = \frac{1}{2\Delta_\tau U} \ln \frac{\cosh{\alpha_i (1+m_i)}}{\cosh{\alpha_i (1-m_i)}}, \label{mtilde_spin} \\
  && C_i = \e^{\Delta_\tau U\tilde{m}^2/2}\cosh{\alpha_i m_i} \label{C_spin}.
\end{eqnarray}
Therefore, if say $m_i$ is given,
$\alpha_i$, $\tilde{m}_i$, and $C_i$ 
are determined from Eqs.~(\ref{alpha_spin})-(\ref{C_spin}). 
Finally we obtain 
\begin{equation}
  \e^{-\Delta_\tau \hat{V}} \propto \prod_i \sum_{s_i=\pm 1}
  \e^{
    \left[ \alpha_i s_i -\frac{\Delta_\tau U}{2} (1-2\tilde{m}_i)\right] \hat{n}_{i\up}
  + \left[-\alpha_i s_i -\frac{\Delta_\tau U}{2} (1+2\tilde{m}_i)\right] \hat{n}_{i\dn}
  -\alpha s_{i} m_i}, 
    \label{propUspin2}
\end{equation}
Note that in general $m_{i} \not = \tilde{m}_i$ and 
$C_i$'s are irrelevant for results of simulations because they cancel out 
from the numerator and the denominator in Eq.~(\ref{afqmc}). 
If $m_i=0$,
the HST reduces to the one introduced by Hirsch~\cite{Hirsch1983}. 
However, the arbitrariness of $m_i$ can be utilized 
to improve the efficiency of AFQMC simulations as shown in Sec.~\ref{sec:4}.

In the right-hand side of Eq.~(\ref{propUspin2}),
the auxiliary field $\alpha_i s_i$ is shifted by $\Delta_\tau U \tilde{m}_i$ 
as compared to the case of $m_{i}=\tilde{m}_i=0$.
To obtain more physical intuitions for $m_i$, we rewrite  
the exponent of the right-hand side of Eq.~(\ref{propUspin2}) as 
\begin{equation}
  \alpha_i s_i \left(\hat{n}_{i\uparrow} - \hat{n}_{i\downarrow} - m_i \right)
  + \Delta_\tau U \tilde{m}_i \left(\hat{n}_{i\uparrow} - \hat{n}_{i\downarrow}\right)
  - \frac{\Delta_\tau U}{2}   \left(\hat{n}_{i\uparrow} + \hat{n}_{i\downarrow}\right).
  \label{eq:exponent}
\end{equation}
In the first term, 
the auxiliary field $\alpha_i s_{i}$ 
is coupled to the fluctuation of the local magnetization
$(\hat{n}_{i\uparrow} - \hat{n}_{i\downarrow} - m_i)$,
while the shift of the local magnetization by $-m_i$ in the first term
is compensated by the spatially inhomogeneous 
magnetic field $\Delta_\tau U \tilde{m}_i$ in the second term.

We set the parameter $m_{i}$ as the 
local magnetization in the trial wave function 
\begin{equation}
  m_{i} =
  \langle \hat{n}_{i\up}  - \hat{n}_{i\dn} \rangle_0.
  \label{eq:mi}
\end{equation}
This $m_i$ can be easily calculated and is expected to stabilize the simulation by keeping 
the first term in Eq.~(\ref{eq:exponent}) ``small'' during the imaginary-time evolution. 
For a given $m_i$, 
$\alpha_i$ can be determined from Eq~(\ref{alpha_spin}),  
$\tilde{m}_i$ from Eq.~(\ref{mtilde_spin}), and  
$C_i$ from Eq.~(\ref{C_spin}). 
The solution $\alpha_i$ of Eq.~(\ref{alpha_spin}) can be found 
by the Newton method with an initial guess 
$\alpha_{i,{\rm initial}}=\cosh^{-1} \e^{\Delta_\tau U/2}$, for example.

\subsection{Auxiliary field coupled to charge fluctuation}\label{hst:charge}
In this subsection, $\alpha_i$ and $C_i$ will be re-defined.
The Hubbard interaction in Eq.~(\ref{HamU}) can be written as 
\begin{eqnarray}
  \hat{V}  
  &=&\frac{U}{2} \sum_{i} 
  \left[\left(\hat{n}_{i\up} + \hat{n}_{i\dn} - \tilde{n}_{i} \right)^2 
    -\tilde{n}_{i}^2\right]\notag \\
  &-&\frac{U}{2} \sum_{i} \left[ 
    (1 - 2 \tilde{n}_{i}) \hat{n}_{i\up} +
    (1 - 2 \tilde{n}_{i}) \hat{n}_{i\dn} \right]
  \label{eq:HamUcharge}
\end{eqnarray}
where $\tilde{n}_i$ is an arbitrary number.  Then $\e^{-\Delta_\tau V}$ can be written as
\begin{eqnarray}
  \e^{-\Delta_\tau \hat{V}} &=& 
  \e^{-\frac{\Delta_\tau U}{2}\sum_{i} \left[(\hat{n}_{i\up} + \hat{n}_{i\dn} - \tilde{n}_{i})^2 - \tilde{n}_{i}^2\right]} \notag \\
  &\times&
  \e^{ \frac{\Delta_\tau U}{2}\sum_{i} (1-2\tilde{n}_i)\hat{n}_{i\up}}
  \e^{ \frac{\Delta_\tau U}{2}\sum_{i} (1-2\tilde{n}_i)\hat{n}_{i\dn}}.  
  \label{propUcharge}
\end{eqnarray}
Let us consider the first exponential factor in 
the right-hand side of Eq.~(\ref{propUcharge}). 
For each site $i$, we consider the following HST: 
\begin{equation}
  C_i \e^{\frac{\Delta_\tau U}{2}\tilde{n}_i^2}
  \e^{-\frac{\Delta_\tau U}{2}(\hat{n}_{i\up} + \hat{n}_{i\dn} - \tilde{n}_i)^2} = 
  \frac{1}{2} \sum_{s=\pm1} \e^{i \alpha_i s_i (\hat{n}_{i\up} + \hat{n}_{i\dn} - n_i)}, 
  \label{assumption_charge}
\end{equation}
where $s_i=\pm 1$ is the discrete auxiliary field,
and the undetermined four parameters $\alpha_i$, $n_i$, $\tilde{n}_i$ and $C_i$ 
are related through the following three equations 
(see Appendix~\ref{appA} for derivation): 
\begin{eqnarray}
    &&\frac{\cos{\alpha_i(2-n_i)} \cos{\alpha_i n_i}}{\cos^2{\alpha_i(1-n_i)}} = \e^{-\Delta_\tau U}\label{alpha_charge}, \\
    &&\tilde{n}_i = 1 - \frac{1}{2 \Delta_\tau U} \ln \frac{\cos{\alpha_i n_i }}{\cos{\alpha_i (2-n_i)}}\label{ntilde_charge}, \\
    &&C_i = \e^{-\Delta_\tau U \tilde{n}_i^2/2}\cos{\alpha_i n_i} \label{C_charge}.
\end{eqnarray} 
Therefore, if say $n_i$ is given,
$\alpha_i$, $\tilde{n}_i$, and $C_i$ 
are determined from Eqs.~(\ref{alpha_charge})-(\ref{C_charge}). 
Finally we obtain 
\begin{equation}
  \e^{-\Delta_\tau \hat{V}} \propto \prod_i \sum_{s_i=\pm 1}
  \e^{
    \left[i \alpha_i s_i +\frac{\Delta_\tau U}{2} (1-2\tilde{n}_i)\right] \hat{n}_{i\up}
  + \left[i \alpha_i s_i +\frac{\Delta_\tau U}{2} (1-2\tilde{n}_i)\right] \hat{n}_{i\dn}
  -i \alpha s_{i} n_i}, 
    \label{propUcharge2}
\end{equation}
Note that in general $n_{i} \not = \tilde{n}_i$ and 
$C_i$'s are irrelevant for results of simulations because they cancel out 
between the numerator and the denominator in Eq.~(\ref{afqmc}). 
If $n_i=1$,
the HST reduces to the one introduced by Hirsch~\cite{Hirsch1983}. 
However, the arbitrariness of $n_i$ can be utilized 
to improve the efficiency of AFQMC simulations as shown in Sec.~\ref{sec:4}.

In the right-hand side of Eq.~(\ref{propUcharge2}),
the auxiliary field $i \alpha_i s_i$ is shifted by $\Delta_\tau U (1-\tilde{n}_i)$ 
as compared to the case of $n_{i}=\tilde{n}_i=1$.
To obtain more physical intuitions for $n_i$, we rewrite  
the exponent of the right-hand side of Eq.~(\ref{propUcharge2}) as 
\begin{equation}
  i  \alpha_i s_i \left(\hat{n}_{i\uparrow} + \hat{n}_{i\downarrow} - n_i \right)
  + \Delta_\tau U (1-\tilde{n}_i) \left(\hat{n}_{i\uparrow} + \hat{n}_{i\downarrow}\right)
  - \frac{\Delta_\tau U}{2} \left(\hat{n}_{i\uparrow} + \hat{n}_{i\downarrow}\right).
  \label{eq:exponent_charge}
\end{equation}
In the first term, 
the auxiliary field $ i \alpha_i s_{i}$ 
is coupled to the fluctuation of the local density
$(\hat{n}_{i\uparrow} + \hat{n}_{i\downarrow} - n_i)$,
while the shift of the local density by $-(1-n_i)$ in the first term
is compensated by the spatially inhomogeneous 
chemical potential $\Delta_\tau U (1-\tilde{n}_i)$ in the second term.

We set the parameter $n_{i}$ as the
local charge density in the trial wave function 
\begin{equation}
  n_{i}
  = \langle \hat{n}_{i\up}  + \hat{n}_{i\dn} \rangle_0. 
  \label{eq:ni}
\end{equation}
This $n_i$ can be easily calculated and is expected to stabilize the simulation by keeping 
the first term in Eq.~(\ref{eq:exponent_charge}) ``small'' during the imaginary-time evolution. 
For a given $n_i$, 
$\alpha_i$ can be determined from Eq~(\ref{alpha_charge}),  
$\tilde{n}_i$ from Eq.~(\ref{ntilde_charge}), and  
$C_i$ from Eq.~(\ref{C_charge}). 
The solution $\alpha_i$ of Eq.~(\ref{alpha_charge}) can be found 
by the Newton method with an initial guess 
$\alpha_{i,{\rm initial}}= \cos^{-1} \e^{-\Delta_\tau U/2}$, for example.

\section{Numerical results}
\label{sec:4}

\subsection{Finite dopings}
At finite dopings, the sign problem occurs~\cite{Hirsch1985,Loh1990}. 
In the presence of the sign problem,
the projection time $\beta$ cannot be taken as large as 
that for the half filled case because the average sign
(of the statistical weight) decreases exponentially in $\beta$~\cite{Loh1990},
otherwise the number of statistical samplings 
has be increased exponentially to keep the statistical error small. 
We set the maximum $\beta$ 
at which the average sign is $\sim 0.05$. 
It will be shown that
even in the presence of the sign problem, 
the AFQMC can still provide 
a good upper bound of the ground-state energy.

\begin{figure}
  \begin{center}
    \includegraphics[width=.95\columnwidth]{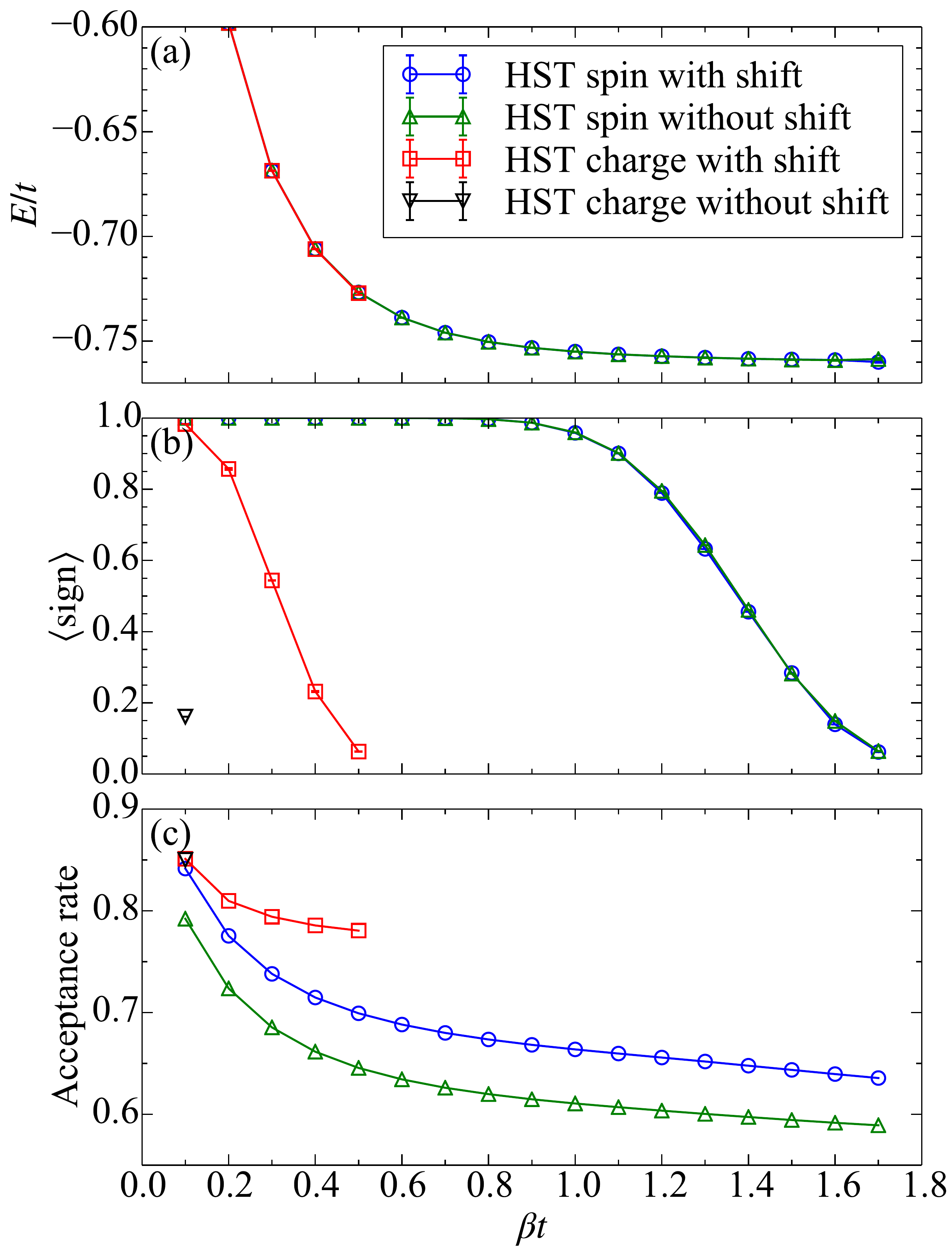}
    \caption{
      \label{fig:energyU8}
      (a) The energy per site, (b) the average sign, and (c) the acceptance rate 
      as a function of the projection time $\beta$ with different HSTs.
      Calculations are done on the $16 \times 16$ cluster with $224$ electrons ($\delta=0.125$)
      at $U/t=8$. 
    }
  \end{center}
\end{figure}

Figure~\ref{fig:energyU8} shows the energy per site 
$E(\beta) = \langle \hat{H} \rangle_\beta /N$, 
the average sign, and the acceptance rate
as a function of $\beta$ at $U/t=8$ 
for the $16 \times 16$ cluster with 
$224$ electrons, corresponding to $\delta = 1/8 = 0.125$.
Note that since 
\begin{equation}
  \frac{\dd E(\beta)}{\dd \beta} = -
  \frac{1}{N}
  \left( \langle \hat{H}^2 \rangle_\beta - \langle \hat{H} \rangle^2_\beta \right) \leqslant 0, 
  \label{eq:slope}
\end{equation}
$E(\beta)$ is a decreasing function of $\beta$ and
its slope $\dd E(\beta)/\dd \beta$ is proportional
to the energy variance~\cite{Horn1984}. 
The energies calculated by different HSTs coincide within the statistical errors,
though their reachable $\beta$ is different, as shown in Fig.~\ref{fig:energyU8}(a).  
In Fig.~\ref{fig:energyU8}(b),
the acceptance rate of the real auxiliary field with the shift (HST spin with shift)
is increased the one from without the shift (HST spin without shift). 
The reason can be attributed to that since 
the first term of Eq.~(\ref{eq:exponent}) with a relevant $m_{i}$
is expected to be ``smaller'' than that with $m_{i}=0$, 
the factor $\e^{\pm 2 \alpha_i (\hat{n}_{i\uparrow} - \hat{n}_{i\downarrow} - m_{i})}$ is closer to unity, 
so that the fluctuation of the norm of the determinant ratio is stabilized.  
On the other hand, the shift of the real auxiliary field
does not affect the average sign significantly 
because the shift does not affect the sign of the determinant ratio, as can be seen in 
Fig.~\ref{fig:energyU8}(c). 
The situation is different for the pure-imaginary auxiliary fields. 
Without the shift (HST charge without shift), 
the average sign diminishes significantly, even at $\beta t=0.1$.
By introducing the shift (HST charge with shift), the average sign
is improved significantly.
The reason can be attributed to that since 
the first term of Eq.~(\ref{eq:exponent_charge}) with a relevant $n_{i}$ 
is expected to be ``smaller'' than that with $n_{i}=1$, 
the factor $\e^{\pm 2 i \alpha_i (\hat{n}_{i\uparrow} + \hat{n}_{i\downarrow} - n_{i})}$ is closer to unity, 
so that the fluctuation of the phase of the determinant ratio is stabilized.  
However, the average sign is still quite smaller 
than that with the real auxiliary fields.
Although the acceptance rate is higher than the real auxiliary fields,
the pure-imaginary fields may not be practical in the presence of the sign problem.  

\begin{figure}
  \begin{center}
    \includegraphics[width=.95\columnwidth]{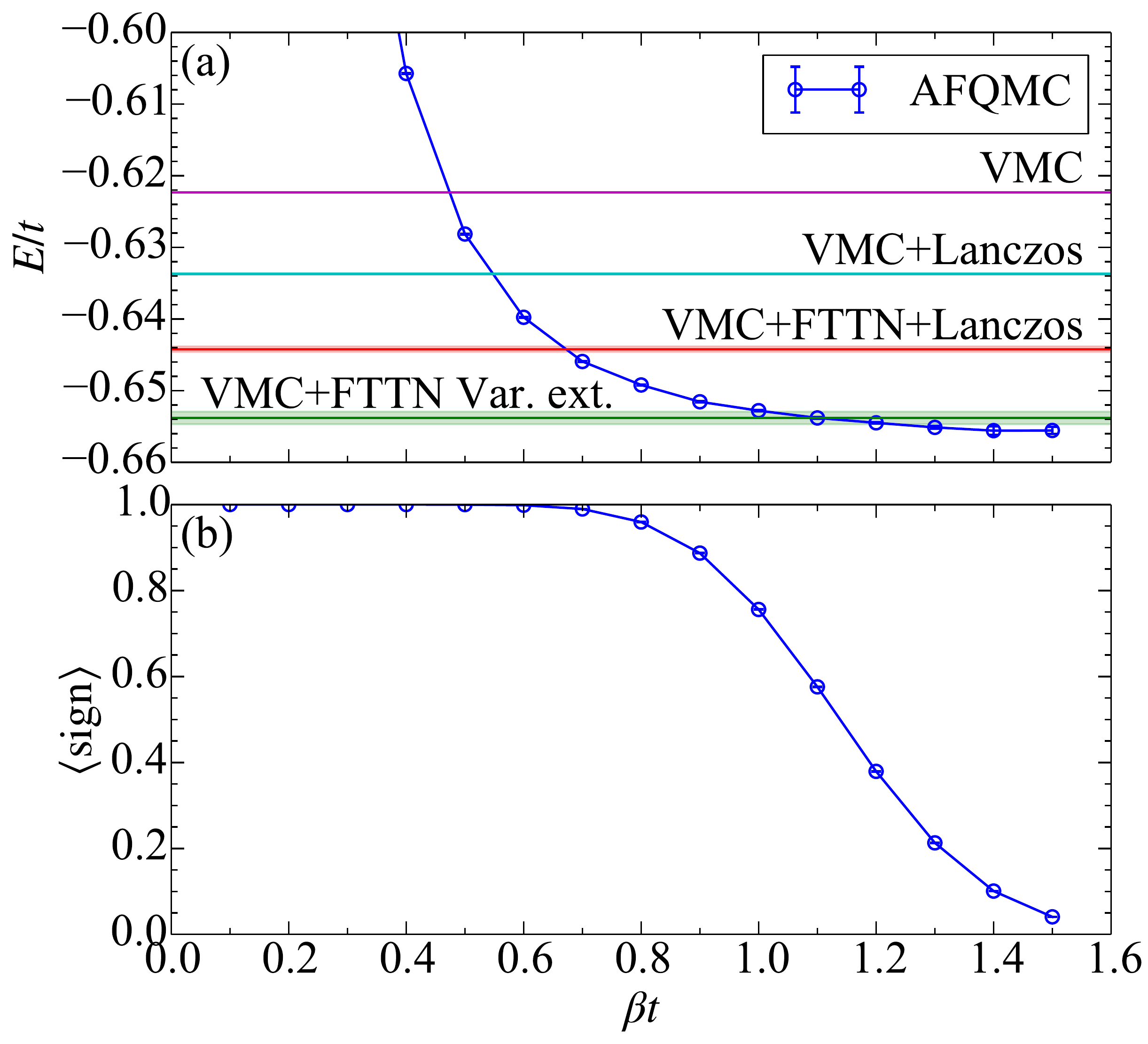}
    \caption{
      \label{fig:energy}
      (a) The energy per site and (b) the average sign
      as a function of the projection time $\beta$.
      Calculations refer to the $16 \times 16$ cluster with $228$ electrons
      ($\delta = 0.109375$) at $U/t=10$. 
      In (a), 
      the horizontal lines and the shaded regions       
      are the VMC energies and their error bars taken from Refs.~\cite{Zhao2017,Ido2018}. 
      FTTN stands for fat-tree tensor network and Var. ext. for variance extrapolation. 
    }
  \end{center}
\end{figure}

To show the usefulness of the AFQMC with a short imaginary-time propagation, 
we make a comparison with the state-of-the-art
variational wave functions for the Hubbard model~\cite{Zhao2017,Ido2018}.
To this end, we move to the smaller doping with the larger $U/t$, 
where the more severe sign problem is expected. 
Figure~\ref{fig:energy} shows the energy per 
and the average sign as a function of $\beta$ at $U/t=10$ 
for the $16 \times 16$ cluster with 
$228$ electrons, corresponding to $\delta =  0.109375$.
Here, only the shifted real auxiliary field is employed because
it turned out to be the most efficient, as shown in Fig.~\ref{fig:energyU8} for $U/t=8$ and $\delta=0.125$.
We use  periodic- (antiperiodic-) boundary condition in the $x$ ($y$) direction
to compare directly with the reference VMC results~\cite{Zhao2017,Ido2018}. 
Notice that our AFQMC energy, computed at finite projection time $\beta$ when the average sign is sufficiently large,  
respects the Ritz's variational principle [see Eq.~(\ref{afqmc})], because it corresponds to the  variational 
expectation value of $\hat{H}$ over the state
$\e^{-\frac{\beta}{2} \hat{H}} | \Psi_{\rm T} \rangle/
\langle \Psi_{\rm T}| \e^{-\beta \hat{H}} | \Psi_{\rm T}\rangle^{\frac{1}{2}}$.
This is a useful property of an approximate technique that is not always satisfied, as for instance 
for the constrained-path AFQMC~\cite{Zhang1997,Zhang2003}.
At $\beta t=0.7$, where the average sign remains $\sim 0.99$,
the AFQMC energy is already lower than 
the VMC energy without variance extrapolation.  
At $\beta t = 1.1$, 
the AFQMC energy almost coincides with the VMC variance-extrapolated one,
while the slope $\dd E(\beta)/\dd \beta$ is still finite, 
indicating that the AFQMC energy variance is nonzero [see Eq.~(\ref{eq:slope})]. 
At $\beta t = 1.5$, 
the AFQMC energy is $E/t=-0.6552(4)$, which is 
lower than the variance-extrapolated VMC energy $E/t=0.6538(9)$~\cite{Zhao2017}
which may be compatible with our number within two standard deviations.
This result suggest that the ground-state AFQMC method 
remains very useful for providing  upper bound values of the ground-state energy 
even in the presence of the negative-sign problem.

\subsection{Half filling}\label{sec:result_half}
At half filling, the sign problem is absent. 
Therefore the AFQMC can provide exact results which often serve as 
a reference benchmark for other numerical techniques.
An excellent agreement in the ground-state energies of 
the two-dimensional Hubbard model between the AFQMC and 
other many-body techniques has been reported in Ref.~\cite{LeBlanc2015}. 
Moreover, within the AFQMC,
the staggered magnetization $m$, i.e., the order parameter
at half filling, can be estimated accurately by using
the twist-averaged boundary condition
for small $U/t$, e.g.,  $U\lesssim 4$~\cite{Qin2016twist,Karakuzu2018twist}.
However, for large $U/t$, 
AFQMC simulations still face a difficulty of 
 large fluctuations of the magnetization, which often lead
to a relatively large error bar in $m$~\cite{Qin2016twist,LeBlanc2015}.
The same difficulty arises also in finite-temperature 
determinant QMC simulations~\cite{Hirsch1989,Varney2009}.
In previous works, in order to overcome the difficulty, a pinning-field method has been  
proposed with a clear improvement for the determination of $m$ in the thermodynamic limit~\cite{Assaad2013,Wang2014}. 
In the following, we report an accurate estimate 
$m$  especially for large $U/t$
by making use of a symmetry-broken trial wave function. 

\begin{figure}
  \begin{center}
    \includegraphics[width=.95\columnwidth]{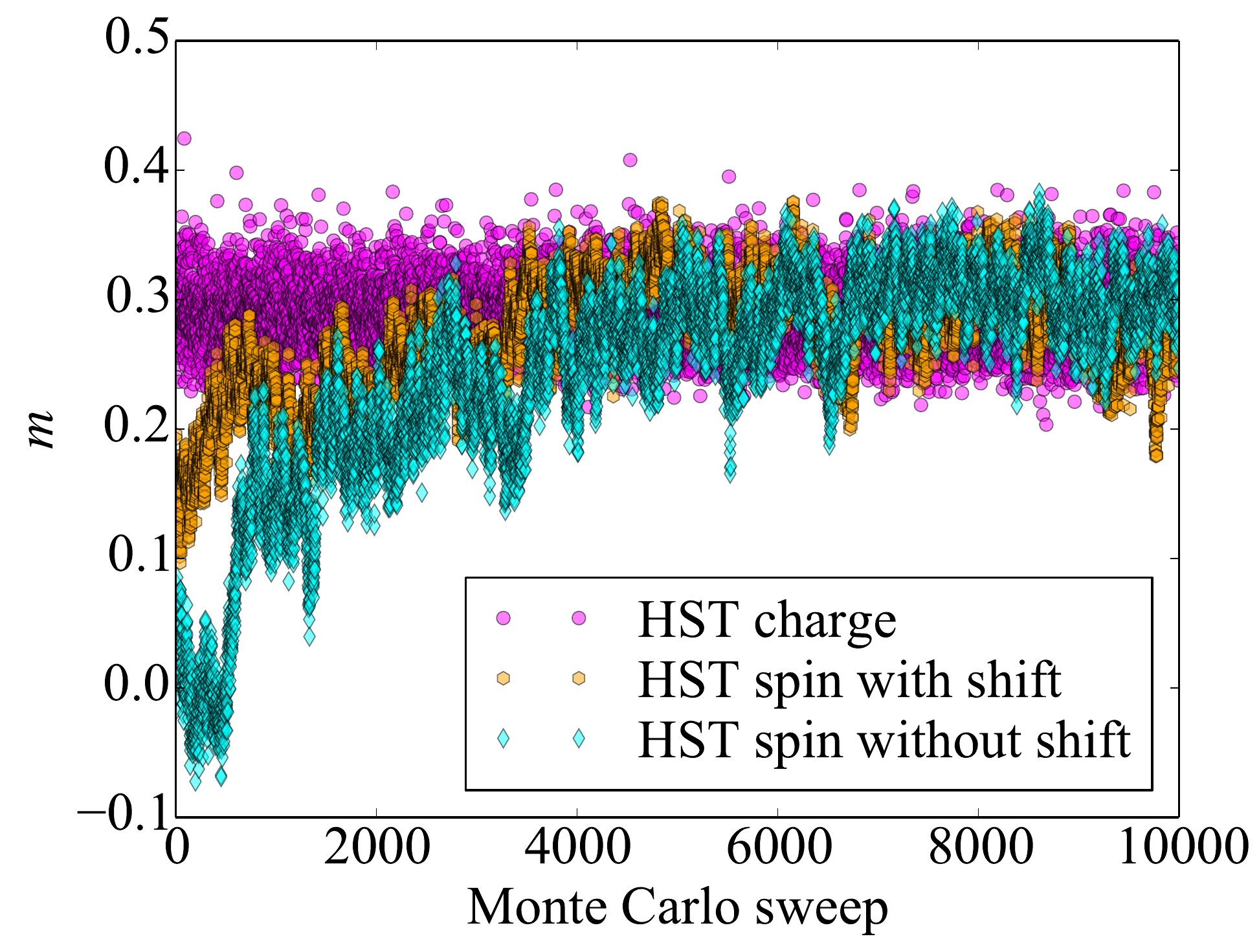}
    \caption{
      \label{fig:maf}
      The staggered magnetization $m$ as a function of the Monte-Carlo sweep for
      the half-filled Hubbard model on the square lattice at $U/t=8$ with different HSTs. 
      Calculations are done on the $16 \times 16$ cluster with $\beta t = 24$. 
    }
  \end{center}
\end{figure}

Figure~\ref{fig:maf} shows the staggered magnetization along the $z$ direction 
\begin{equation}
  m(\beta) = \frac{1}{2N} \sum_{i} (-1)^i
  \langle \hat{n}_{i \up} - \hat{n}_{i \dn} \rangle_\beta
  \label{eq:mag}
\end{equation}
as a function of the Monte Carlo sweep with different HSTs.
The calculations are done for $U/t=8$, $\beta t=24$, and $\Delta_\tau t = 0.1$ 
on the $L=16$ cluster with periodic-boundary conditions. 
We use $\Delta_{\rm AF}/t=0.001$ to give a finite staggered magnetization
in the trial wave function. This small value of $\Delta_{\rm AF}$ is effective to pin a sizable 
value of the finite size order parameter $m(0)$ because the single-particle states at $U/t=0$
have a large degeneracy ($\propto L$) at the Fermi level,and are therefore  strongly renormalized upon an arbitrary small 
$\Delta_{\rm AF}$.

Since $m(0)$ is finite [see Eq.~(\ref{eq:Hmf})], 
$m(\beta)$ remains finite even for finite $L$. 
Note that, at half filling, 
the HST in the charge channel with shift is equivalent to
the one without shift because $n_i=1$. 
A very large equilibration time of $\sim 5000$ Monte Carlo sweeps is found
for $m$ with the standard real HST coupled to the on-site  electron spins. 
In this case, our shifted HST improves the equilibration time, allowing also  
higher acceptance rate (not shown) as in the doped cases,
but the improvement is not really important.  
Amazingly, $m$ is equilibrated almost immediately 
for the complex HST coupled to the on-site electron charges.  
This result implies that this pure-imaginary auxiliary field, 
which was first  introduced by Hirsch~\cite{Hirsch1983}, 
is very useful to estimate $m$ at half filling for large $U/t$. 
Here we emphasize also that, not only the correlation time
is highly reduced with this technique, but also fluctuations,
thanks to this pinning strategy in the trial wave function,
do not show any problem of large fluctuations, even at very large $U/t$ and values.

\begin{figure}
  \begin{center}
    \includegraphics[width=.95\columnwidth]{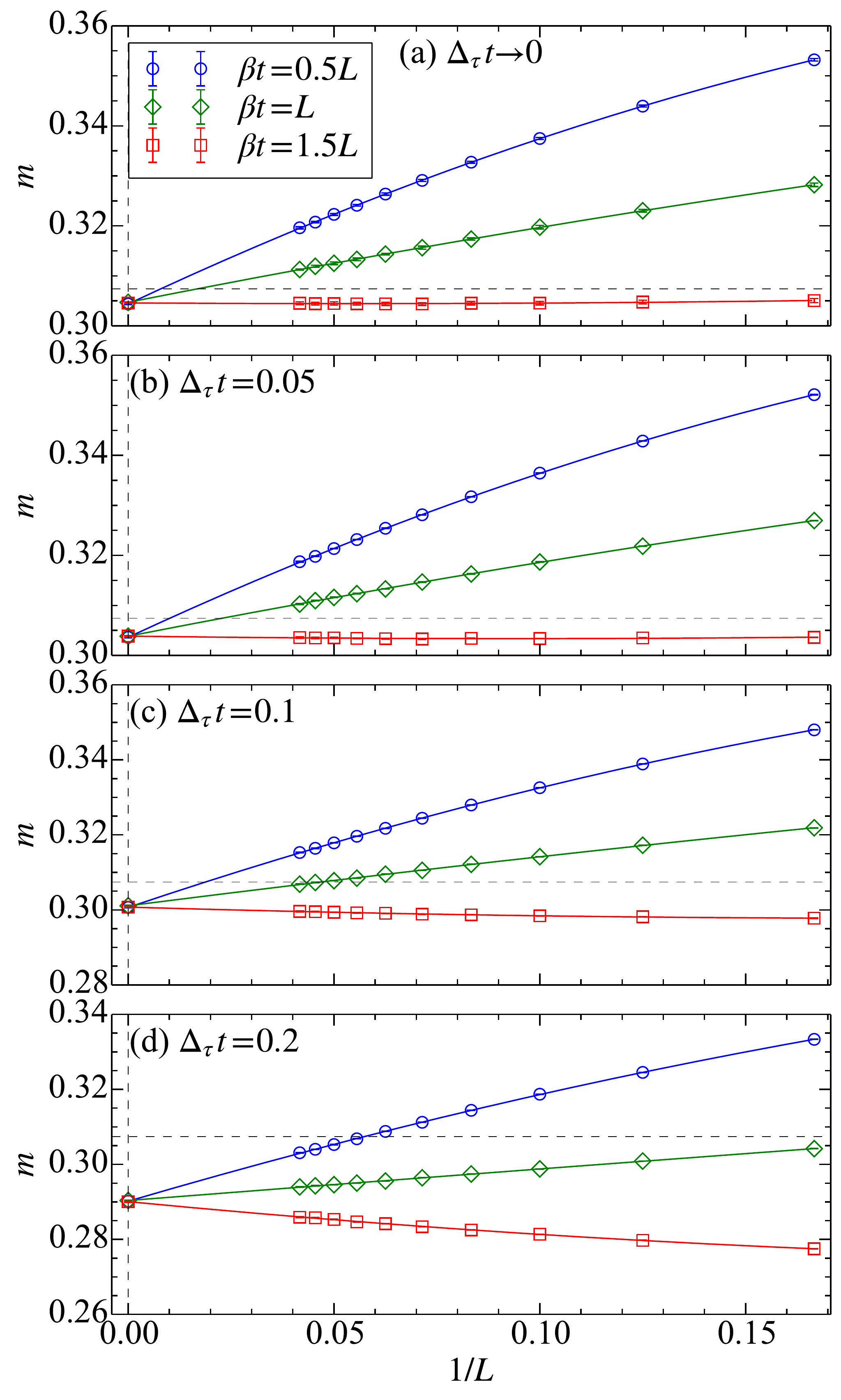}
    \caption{
      \label{fig:mvsl}
      Finite-size scaling of the staggered magnetization $m$
      of the half-filled Hubbard model at $U/t=10$ with
      $\beta t =0.5L, L$, and $1.5L$ for
      (a) $\Delta_\tau t \to 0$, 
      (b) $\Delta_\tau t = 0.05$, 
      (c) $\Delta_\tau t = 0.1$, and 
      (d) $\Delta_\tau t = 0.2$.   
      The dashed horizontal lines indicate $m$ of the Heisenberg model
      in the thermodynamic limit taken from Ref.~\cite{Sandvic2010}.
    }
  \end{center}
\end{figure}

Figure~\ref{fig:mvsl} shows
the finite-size scaling of $m$ for $U/t=10$.
The cluster sizes used are $L=6, 8, 10, 12, 14, 16, 18, 20, 22,$ and $24$. 
Here, the projection time $\beta$ is chosen proportional to $L$, i.e.,
$\beta t = \alpha L$, with $\alpha=0.5, 1$ and $1.5$.
These $\beta t$ values are an order of magnitude smaller than 
those used with the pinning-field method~\cite{Assaad2013,Wang2014}, because in our 
approach we can reach the thermodynamic limit consistently without unnecessarily large values of $\beta$.
Indeed, the extrapolated values at $1/\beta=1/L=0$ are consistent for all values of $\alpha$,
which validates our approach. 
Our best estimate is obtained from the $\beta t = 1.5L$ set of data, 
yielding $m=0.3046(1)$ in the $\Delta \to 0$ limit, 
where the number in the parentheses indicates the extrapolation error in the last digit. 
Calculations are done for
$\Delta_\tau t = 0.2$, $0.1$, and $0.05$ and the extrapolations 
to $\Delta_\tau \to 0$ are obtained  by a linear 
fit in $(\Delta_\tau t)^2$, determined by the least-squares method.
The ground-state expectation value $m$ in the thermodynamic limit is 
obtained by extrapolating the results to $L\to \infty$. 
In this case we fit the data in the range $6 \leqslant L \leqslant 24$
with quadratic polynomials in $1/L$.
As it can be seen in Fig.~\ref{fig:mvsl}, the time-discretization error
is not negligible for $U/t=10$. 
The extrapolated value is certainly smaller than the one 
in the Heisenberg model~\cite{Anderson1952,Reger1988,Sandvic1997,Calandra1998}, 
where the latest Monte Carlo estimate is $m=0.30743(1)$~\cite{Sandvic2010,Jiang2011}.  

\begin{figure}
  \begin{center}
    \includegraphics[width=.95\columnwidth]{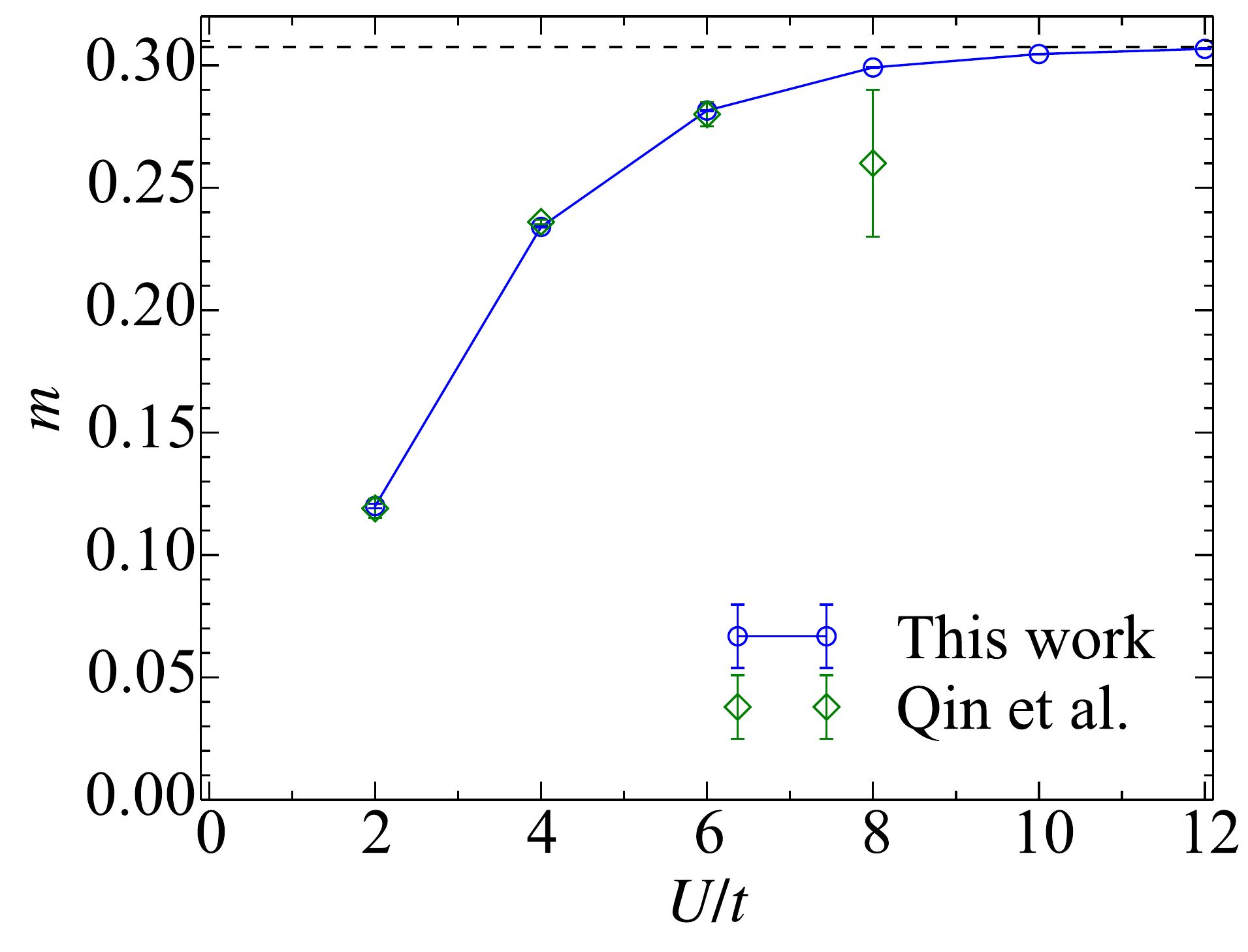}
    \caption{
      \label{fig:mvsU}
      The staggered magnetization $m$ in the thermodynamic limit as a function of $U$.
      For comparison, previous AFQMC results are taken from Ref.~\cite{Qin2016}. 
      The dashed horizontal line indicates $m$ of the Heisenberg model
      in the thermodynamic limit taken from Ref.~\cite{Sandvic2010}.
    }
  \end{center}
\end{figure}

In Table~\ref{tab:m} and Fig.~\ref{fig:mvsU} we show  the values for $m$ 
in the thermodynamic limit for $U/t=2,4,6,8,10$ and $12$ 
and compare them with the ones available in the
literature~\cite{Karakuzu2018,Sorella2015,Qin2016twist,Karakuzu2018twist}.
The main outcome of this work  is the estimated value of $m$ for $U/t\geqslant 8$, which is usually 
the accepted value for cuprates.
Here, our error bar at $U/t=8$ is two orders of magnitude smaller than the 
previous AFQMC estimate~\cite{LeBlanc2015,Qin2016twist}.
Thanks to this high statistical accuracy,
our results clearly show that $m$ increases monotonically in $U/t$. 
This is consistent with a strong-coupling expansion around the Heisenberg limit~\cite{Delannoy2005}.
Here, finite-size scaling analyses are performed as follows. 
For $U/t \geqslant 6$, the scheme of finite-size scaling analyses 
is the same as that for $U/t = 10$ which has been described before.  
For $U/t=4$ ($U/t=2$), cluster sizes up to $L=32$ ($L=50$) 
with twist-averaged boundary conditions~\cite{gros_exact_diag,gros_gtabc,Koretsune2007,Qin2016twist,Karakuzu2017,Karakuzu2018twist} are used 
because the finite-size effects are more important than
those we have found  at  larger $U/t$ values. 
A much larger value of $\Delta_{\rm AF}/t=10$  is used for $U/t \leqslant 4$ 
because the twists remove the degeneracy of the 
single-particle states at $U/t=0$, as discussed before. 
All the results are obtained in the $\Delta_\tau \to 0$ limit 
using data at 
$\Delta_\tau t = 0.2$, $0.1$, and $0.05$ for $U/t \geqslant 4$ and
$\Delta_\tau t = 0.25$, $0.2$, and $0.1$ for $U/t = 2$, respectively.

\begin{table*}
\caption{The staggered magnetization $m$ of the two-dimensional Hubbard model
  at half filling in the thermodynamic limit. 
  The staggered magnetization of the two-dimensional Heisenberg model 
  from Refs.~\cite{Sandvic2010,Jiang2011} is also shown. 
  PBC stands for periodic boundary condition, 
  TABC for twist-averaged boundary conditions, and 
  MBC for modified boundary condition.
  \label{tab:m}}
\centering
\begin{adjustbox}{width=1\textwidth}
  {
  \begin{tabular}{llllllll}
    \hline
    \hline
    $U/t$                                      & $2$       & $4$       & $6$       & $8$       & $10$      & $12$   & $\infty$ (Heisenberg antiferromagnet) \\ 
    \hline
    AFQMC     (this work)               & 0.120(1)  & 0.2340(2) & 0.2815(2) & 0.2991(2) & 0.3046(1) & 0.3067(2) & --  \\
    AFQMC TABC~\cite{Qin2016twist}      & 0.119(4)  & 0.236(1)  & 0.280(5)  & 0.26(3)   & --        & --            & --  \\
    AFQMC TABC~\cite{Karakuzu2018twist} & 0.122(1)  & 0.2347(4) & --        & --        & --        & --            & --  \\
    AFQMC PBC~\cite{Karakuzu2018}       & --        & 0.238(3)  & --        & --        & --        & --            & --  \\
    AFQMC MBC~\cite{Sorella2015}        & 0.120(5)  & --        & --        & --        & --        & --            & --  \\
    QMC Heisenberg model~\cite{Sandvic2010,Jiang2011}  & --        & --        & --        & --        & --        & --            & 0.30743(1)  \\
    \hline
    \hline
  \end{tabular}
  }
  \end{adjustbox}
\end{table*}

\section{Conclusions and Discussions}
\label{sec:5}
In this work we have shown that, within the ground-state AFQMC technique, the choice of the 
trial function and the one for the auxiliary field are extremely important. 
In particular we have improved the efficiency of the method, by introducing 
shifted-discrete HSTs, that are useful
for performing the imaginary-time evolution of 
 symmetry-broken trial wave functions.
The formalism can be considered as a generalization of the 
discrete HSTs in Ref.~\cite{Hirsch1983} or a compactification of 
the shifted-contour auxiliary-field Monte Carlo formalism in Ref.~\cite{Rom1997,Rom1998}
specialized to the on-site Hubbard interaction. 

Properly chosen auxiliary fields can improve
the efficiency of AFQMC simulations. 
The shifted real auxiliary fields can improve the acceptance ratio, while
the shifted pure-imaginary auxiliary fields can improve the average sign.
The reason is that the shift in the real auxiliary field
can stabilize the fluctuations of the norm of the determinant ratio,
while the shift in the pure-imaginary auxiliary field 
can stabilize the fluctuations of the phase of the determinant ratio.
However, even after the improvement, the average sign with
the pure-imaginary auxiliary field remains worse  than
the one obtained  with real auxiliary field for the doped cases. 
Therefore, in the presence of the sign problem, 
the real auxiliary field is still recommended 
for achieving longer imaginary-time propagations.
On the other hand, at half filling with large $U/t$,
the pure-imaginary auxiliary field is 
dramatically more efficient than the real-auxiliary fields for
evaluating the staggered magnetization $m$. 

In our approach, $m_{i}$ or $n_{i}$ in Eqs.~(\ref{eq:HamUspin}) or (\ref{eq:HamUcharge}) are arbitrary parameters, 
that do not have to be necessarily chosen as 
in Eq.~(\ref{eq:mi}) or in Eq.~(\ref{eq:ni}). 
For example, $m_i$ ($n_i$) can be updated iteratively 
by the AFQMC expectation value of 
$\hat{n}_{i \uparrow} - \hat{n}_{i \downarrow}$ 
($\hat{n}_{i \uparrow} + \hat{n}_{i \downarrow}$) 
with iterative simulations. 
This kind of scheme has already been  employed to construct self-consistently an optimized 
trial wave function in the AFQMC~\cite{Qin2016}.
Obviously, shifted-discrete HSTs can  be used straightforwardly also in this case.  
Moreover, we expect  that 
imaginary-time dependent $m_i$ or $n_i$ 
could further improve the efficiency of the AFQMC, especially within the constrained path formalism. 
A study along this line is in progress~\cite{SorellaAAD}.

Finally, we remark on
the $d$-wave superconducting order 
which has not been considered in the present study. 
It is noteworthy that an early study on a $t$-$t'$-$J$ model~\cite{Himeda2002} 
has shown that a stripe state with spatially oscillating $d$-wave superconductivity
is favored around $1/8$ hole doping. Considering such an inhomogeneous
superconductivity in a trial wave function
might be of interest for a possible improvement 
of AFQMC simulations for doped Hubbard models with large $U/t$. 

\acknowledgements
The authors would like to thank Seher Karakuzu, Federico Becca, 
Luca Fausto Tocchio, and Tomonori Shirakawa for helpful discussions.
K.S. acknowledges Emine K{{\"u}}{{\c{c}}}{{\"u}}kbenli and Stefano de Gironcoli 
for bringing his attention to Refs.~\cite{Perdew1995,Carrascal2015}. 
Computations have been done by using the HOKUSAI GreatWave and HOKUSAI BigWaterfall 
supercomputers at RIKEN under the Projects No.~G18007 and No.~G18025.
K.S. acknowledges support from the JSPS Overseas Research Fellowships. 
S.S. acknowledges support by the Simons foundation.

\clearpage
\appendix
\section{Derivation of shifted-discrete HSTs}
\label{appA}
In this Appendix, we derive
Eqs.~(\ref{alpha_spin})-(\ref{C_spin}) and 
Eqs.~(\ref{alpha_charge})-(\ref{C_charge}). 
First we derive Eqs.~(\ref{alpha_spin})-(\ref{C_spin}), i.e.,
the shifted-discrete HST in the spin channel. 
Since the fermion density operator $\hat{n}_{i\s}$ is idempotent,
i.e., $\hat{n}_{i \s}^2 = \hat{n}_{i \s}$, 
its exponential function is written as 
\begin{equation}
  \e^{\alpha s \hat{n}_{\s}} = 1 + \left(\e^{ \alpha s} - 1\right) \hat{n}_{\s}, 
  \label{idem}
\end{equation} 
where, and hereafter, the site index $i$ is dropped for brevity. 
Then the right-hand side of Eq.~(\ref{assumption}) is given as 
\begin{eqnarray}
  &&
  \frac{1}{2}\sum_{s=\pm1}
  \left[
    1 + \left(\e^{\alpha s} -1 \right) \hat{n}_{\up}
    \right]
  \left[
    1 + \left(\e^{-\alpha s} -1 \right) \hat{n}_{\dn}
    \right]
  \e^{-s \alpha m}\notag \\
  &=&
  \cosh{\alpha m} \notag \\
  &+& \left[\cosh{\alpha(1-m)} - \cosh{\alpha m} \right] \ \hat{n}_\up \notag \\ 
  &+& \left[\cosh{\alpha(1+m)} - \cosh{\alpha m} \right] \ \hat{n}_\dn \notag \\ 
  &+& \left[2\cosh{\alpha m} - \cosh{\alpha(1-m)} - \cosh{\alpha(1+m)} \right] \hat{n}_\up \hat{n}_\dn. 
  \label{hst_right}
\end{eqnarray}
The left-hand side of Eq.~(\ref{assumption}) is given as 
\begin{eqnarray}
  &&C \e^{- \Delta_\tau U \tilde{m}^2/2}
  \e^{\frac{\Delta_\tau U}{2}(1-2\tilde{m}) \hat{n}_\up} 
  \e^{\frac{\Delta_\tau U}{2}(1+2\tilde{m}) \hat{n}_\dn}   
  \e^{-\Delta_\tau U \hat{n}_\up \hat{n}_\dn} \notag \\
  &=& C \e^{- \Delta_\tau U \tilde{m}^2/2} \notag \\
  &+& C \e^{- \Delta_\tau U \tilde{m}^2/2} \left[\e^{\frac{\Delta_\tau U}{2}(1-2\tilde{m})}-1 \right] \hat{n}_\up \notag \\
  &+& C \e^{- \Delta_\tau U \tilde{m}^2/2} \left[\e^{\frac{\Delta_\tau U}{2}(1+2\tilde{m})}-1 \right] \hat{n}_\dn \notag \\
  &+& C \e^{- \Delta_\tau U \tilde{m}^2/2} \left[2 
  - \e^{\frac{\Delta_\tau U}{2}(1-2\tilde{m})} 
  - \e^{\frac{\Delta_\tau U}{2}(1+2\tilde{m})} \right]
  \hat{n}_\up \hat{n}_\dn.
  \label{hst_left}
\end{eqnarray}
By comparing Eq.~(\ref{hst_right}) with Eq.~(\ref{hst_left}), we obtain   
Eqs.~(\ref{alpha_spin})-(\ref{C_spin}).  

Next, we derive Eqs.~(\ref{alpha_charge})-(\ref{C_charge}), i.e.,
the shifted-discrete HST in the charge channel. 
As in Eq.~(\ref{idem}), we have 
\begin{equation}
  \e^{i \alpha s \hat{n}_{\s}} = 1 + \left(\e^{i \alpha s} - 1\right) \hat{n}_{\s}.
  \label{idem:charge}
\end{equation} 
Then the right-hand side of Eq.~(\ref{assumption_charge}) is given as 
\begin{eqnarray}
 &&
  \frac{1}{2}\sum_{s=\pm1}
  \left[
    1 + \left(\e^{i \alpha s} -1 \right) \hat{n}_{\up}
    \right]
  \left[
    1 + \left(\e^{i \alpha s} -1 \right) \hat{n}_{\dn}
    \right]
  \e^{-i s \alpha n}\notag \\
  &=&
  \cos{\alpha n} \notag \\ &+&
  \left[\cos{\alpha(1-n)} - \cos{\alpha n})\right]
  (\hat{n}_\up + \hat{n}_\dn) \notag \\
  &+& \left[\cos{\alpha(2-n)} - 2\cos{\alpha(1-n)} + \cos{\alpha n} \right] \hat{n}_\up \hat{n}_\dn. 
  \label{hstc_right}
\end{eqnarray}
The left-hand side of Eq.~(\ref{assumption_charge}) is given as 
\begin{eqnarray}
  &&
  C
  \e^{\Delta_\tau U \tilde{n}^2/2}
  \e^{-\frac{\Delta_\tau U}{2}(1-2\tilde{n})(\hat{n}_\up + \hat{n}_\dn)}
  \e^{-\Delta_\tau U \hat{n}_\up \hat{n}_\dn} \notag \\
  &=&
  C\e^{\Delta_\tau U \tilde{n}^2/2} \notag \\ &+&
  C\e^{\Delta_\tau U \tilde{n}^2/2}
  \left[\e^{-\frac{\Delta_\tau U}{2}(1-2\tilde{n})}-1 \right]
  \left( \hat{n}_\up  + \hat{n}_\dn \right) \notag \\ &+&
  C\e^{\Delta_\tau U \tilde{n}^2/2} \left[
    \e^{-\Delta_\tau U}
    \e^{-\Delta_\tau U(1-2\tilde{n})}
    - 2 \e^{-\frac{\Delta_\tau U}{2}(1-2\tilde{n})} + 1\right] \hat{n}_\up \hat{n}_\dn.
  \label{hstc_left}
\end{eqnarray}
By comparing Eq.~(\ref{hstc_right}) with Eq.~(\ref{hstc_left}), we obtain   
Eqs.~(\ref{alpha_charge})-(\ref{C_charge}).  

\bibliography{biball}
\end{document}